\documentclass[journal]{IEEEtran}
\usepackage{lipsum}
\makeatletter
\def\footnoterule{\relax
	\kern-5pt
	\hbox to \columnwidth{\hfill\vrule width 0.5\columnwidth height 0.4pt\hfill}
	\kern4.6pt}
\makeatother
\usepackage{amsfonts}
\usepackage{amssymb}
\usepackage{latexsym}
\usepackage{cite}
\usepackage[tight,footnotesize]{subfigure}
\usepackage{xcolor}
\usepackage{mathtools}
\usepackage{amsmath}

\usepackage[utf8]{inputenc}
\usepackage[autostyle]{csquotes}
\usepackage{textcomp}

\usepackage{siunitx}
\usepackage{epstopdf}
\usepackage{graphicx,epsfig}

\usepackage{todonotes}

\usepackage[inline]{enumitem}

\definecolor{morange}{rgb}{0.8,0.4,0.1}
\definecolor{mblue}{rgb}{0,0,0.9}
\definecolor{mdblue}{rgb}{0,0.25,0.5}
\definecolor{mred}{rgb}{0.8,0.1,0}
\definecolor{mgreen}{rgb}{0.1,0.6,0}

\newcommand{\mm}{\si{\milli\meter}}
\newcommand{\onebyone}{$1\times1$ }
\newcommand{\twobytwo}{$2\times2$ }
\newcommand{\onebyfour}{$1\times4$ }
\newcommand{\fourbyfour}{$4\times4$ }
\newcommand{\nbyn}{$N\times N$ }
\newcommand{\onebyn}{$1\times N$ }

\newcommand{\Sparam}{${S}$-parameter}
\newcommand{\apo}{\textquotesingle}

\newcommand{\etal}{\textit{et al. }}

\usepackage[belowskip=-10pt]{caption}
\begin{document}


\title{RF Lens-Embedded Antenna Array for \\mmWave MIMO: Design and Performance}

\author{Yae Jee Cho,~\IEEEmembership{Student~Member,~IEEE,}
	Gee-Yong Suk,~\IEEEmembership{Student~Member,~IEEE,} Byoungnam~Kim,~\IEEEmembership{Member,~IEEE,}
	Dong Ku~Kim~\IEEEmembership{Senior~Member,~IEEE},
	 and Chan-Byoung~Chae,~\IEEEmembership{Senior~Member,~IEEE}

\thanks{Y. J. Cho, G. -Y. Suk, and C.-B. Chae are with the School of Integrated Technology, Yonsei University, Korea (e-mail: {yjenncho, gysuk, cbchae}@yonsei.ac.kr). B. Kim is with SENSORVIEW, Korea (e-mail: Klaus.kim@sensor-view.com). D. K. Kim is with the School of Electrical and Electronic Engineering, Yonsei University, Korea (email: dkkim@yonsei.ac.kr). Corresponding author is C.-B. Chae.}
\thanks{This research was supported by Institute for Information \& Communications Technology Promotion (IITP) grant funded by the Korea Government (MSIT) (No. 2016-0-00208, High Accurate Positioning Enabled MIMO Transmission and Network Technologies for Next 5G-V2X Services).}
}

\maketitle

\begin{abstract}
The requirement of high data-rate in the fifth generation wireless systems (5G) calls for the ultimate utilization of the wide bandwidth in the mmWave frequency band. Researchers seeking to compensate for mmWave\apo s high path loss and to achieve both gain and directivity have proposed that mmWave multiple-input multiple-output (MIMO) systems make use of beamforming systems. Hybrid beamforming in mmWave demonstrates promising performance in achieving high gain and directivity by using phase shifters at the analog processing block. What remains a problem, however, is the actual implementation of mmWave beamforming systems; to fabricate such a system is costly and complex. With the aim of reducing such cost and complexity, this article presents actual prototypes of the lens antenna as an effective device to be used in the future 5G mmWave hybrid beamforming systems. Using a lens as a passive phase shifter enables beamforming without the heavy network of active phase shifters, while gain and directivity are achieved by the energy-focusing property of the lens. Proposed in this article are two types of lens antennas, one for static and the other for mobile usage. Their performance is evaluated using measurements and simulation data along with link-level analysis via a software defined radio (SDR) platform. Results show the promising potential of the lens antenna for its high gain and directivity, and its improved beam-switching feasibility compared to when a lens is not used. System-level evaluations reveal the significant throughput enhancement in both real indoor and outdoor environments. Moreover, the lens antenna\apo s design issues are also discussed by evaluating different lens sizes. 
\end{abstract}

\begin{IEEEkeywords}
mmWave communications, beamforming, lens antenna prototype, design, and performance.
\end{IEEEkeywords}



\section{Introduction}
\IEEEPARstart{F}{or} next generation wireless applications, the required data-rate is increasing tremendously. The specified requirement for low-mobility users in fifth generation wireless systems (5G) is at least 10--50~Gbps~\cite{AOss20145Greq}, 10 times the Long Term Evolution-Advanced (LTE-A) peak data-rate. Clearly, the currently used frequency--sub-\SI{6}{\giga\hertz}--is too limited and crowded for 5G utilization. An intuitive approach to achieve higher data-rate would be to utilize the large frequency resources available at the mmWave range (30--\SI{300}{\giga\hertz}). LTE-A Pro, with sub-\SI{6}{\giga\hertz} center frequency, has an average bandwidth of \SI{100}{\mega\hertz}~\cite{LTEWhitePaper_Nokia} while the currently operating mmWave based standard {IEEE} 802.15.3.c-2009 has 20 times that bandwidth (2--\SI{9}{\giga\hertz}). This wide bandwidth potential has attracted much research on mmWave wireless communications. 

One of the most studied aspects of utilizing mmWave is how to overcome the high free-space path loss. The high path loss causes severe attenuation in the signal when confronting blockages or being transmitted across long distances. Previous research presented solutions to this problem by designing a small but compact massive multiple-input multiple-output (MIMO) using the short wavelength of mmWave. A much more compact massive MIMO with more antenna elements and higher gain can be made compared to what is achievable at the sub-\SI{6}{\giga\hertz} frequency range. This larger antenna array gain over-compensates mmWave\apo s~path loss.

\begin{figure*}[t]
	\centering{\includegraphics[width=1.8\columnwidth,keepaspectratio]
		{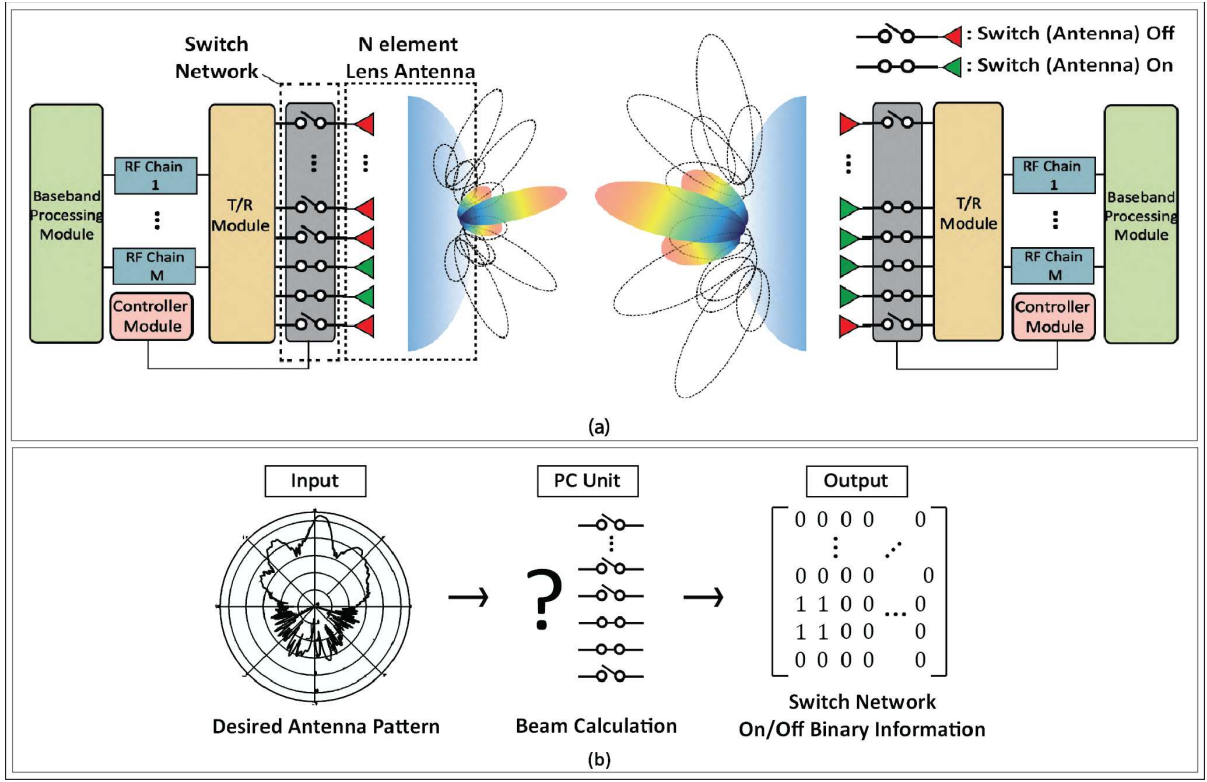}}
	\caption{(a) Proposed mmWave lens hybrid beamforming system; (b) a schematic of the controller module used in the proposed mmWave lens hybrid beamforming system.} \label{fig1:lensbf_archi}
\end{figure*}

To further increase the array gain of the mmWave massive MIMO, the signal phase for each antenna element can be shifted by additional processing modules such as phase shifters. This results in an antenna radiation pattern of high directivity and gain towards a certain predetermined direction. Applying phase shifting to each of the multiple data streams after the baseband to form highly directed beams in multiple directions is called hybrid beamforming. The beamforming gain is, in general, proportional to the antenna array size and dimension, which is an advantage delivered by the compactness of the mmWave massive MIMO. 

\begin{figure*}[t]
	\centering{\includegraphics[width=1.7\columnwidth,keepaspectratio]{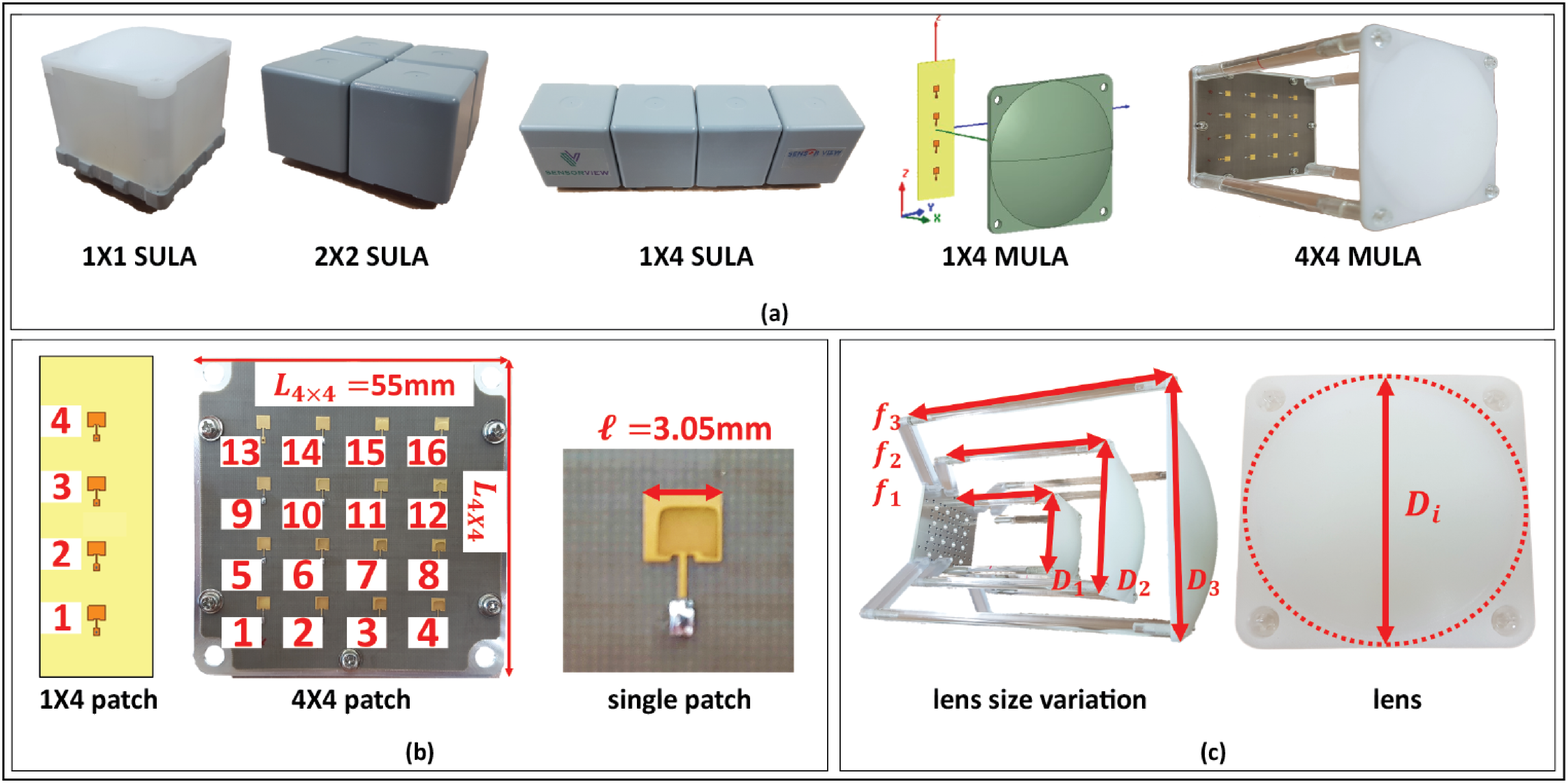}}
	\caption{SULA and MULA prototypes and fabrication details: (a) fabricated prototypes of the SULA and the MULA; (b) details of the fabricated patch antenna array and the assigned port numbers; (c) type of lens and size parameters.} \label{fig2:antenna_pic}
\end{figure*}
In prior work, prototypes for beamforming using phase shifters~\cite{hbf_samsung,Jang_WCM16} or Butler matrix~\cite{Chang_But2010} were presented to demonstrate the feasibility of mmWave beamforming. While the Butler matrix is simple and easy to realize, its limited, low achievable gain and scanning plane make it hard to utilize for actual 5G applications. Regarding phase shifters, in \cite{hbf_samsung}, Roh \etal presented a \SI{27.9}{\giga\hertz} operating analog beamfomring prototype using $8\times4$ uniform planar arrays (UPAs) at the transceivers with \SI{520}{\mega\hertz} bandwidth orthogonal frequency-division multiplexing (OFDM) settings. In non-line-of-sight (NLoS) outdoor environments, the prototype supported a data-rate of more than 500 Mbps with 8~km/h mobility. 
In \cite{Jang_WCM16}, Jang and colleagues also examined the feasibility of three-dimensional (3D) hybrid beamforming with phase shifters. The authors proposed an algorithm for beamforming without channel state information (CSI) and tested it using a real-time testbed with software-defined radios (SDR) and fabricated dual-pole antenna arrays. Under LTE environments, the proposed system gave higher average data-rate of 32~Mbps compared to the conventional LTE system in indoors.

The aforementioned studies all demonstrated through actual prototypes that beamforming is indeed feasible and significant. However, the crucial drawback of implementing beamforming via phase shifters is that the prototypes had to compromise their number of radio frequency (RF) chains or antenna array size due to the high power consumption and hardware complexity involved in implementing the actual network of analog-to-digital converters (ADCs) and phase shifters. Considering the large number of antenna elements in a mmWave massive MIMO, such drawbacks of complexity and power consumption should be dealt with to actualize 5G mmWave hybrid beamforming.
The research in \cite{akbar2013beam,RuiZhang2016opdm,sayeed2013MUbeamspace,al2011sppLens} incorporated a lens in place of the conventional phase shifters into the beamforming architecture to reduce computational complexity and power consumption. The lens acts as a virtual passive phase shifter, focusing the incident electromagnetic wave to a certain region. This lens, when used jointly with antennas, exhibits two significant properties:  
\textit{\begin{enumerate*}[label=(\roman*),before=\unskip{~}, itemjoin={{, and }}] 
	\item focused signal power at the front end achieving high directivity and gain\footnote{Although the concept of gain and directivity are correlated, we use the two terms distinctly throughout the article. Gain means the maximum achievable gain of the antenna, and directivity means the capacity of the antenna to generate a sharp beam.}
	\item concentrated signal power directed to a sub-region of the antenna array. 
\end{enumerate*}}
These properties make the lens a practical tool for implementing the RF front-end in beamforming systems. It not only enables improvements in system performance by increasing gain and directivity via energy focusing, but also reduces signal processing complexity and RF chain cost by allowing only a subset of antenna elements to be activated instead of all. 

In this article, we evaluate the properties and feasibility of mmWave lens antennas aimed to be used in hybrid beamforming structures. We use actual lens antenna prototypes that differ in design, and compare them with the case when lens is not used. The lens antenna is designed differently according to whether it is employed in static or mobile communication links. Regarding actual low-cost, low-complexity beamforming lens modules, to the best of our knowledge, there is no prior research that analyzes and compares the performance of fabricated mmWave lens modules with different structures and to cases in which a lens is not used through link- and system-level evaluations. 

In Section~II, we elaborate on the general concept of a lens antenna used in beamforming systems, and present the relevant previous research along with a schematic of our aimed mmWave lens hybrid beamforming system. Section III describes the details of our proposed lens antenna prototypes, including measurement results and analysis on gain, \Sparam s, and radiation patterns. Link-level evaluation is also presented. In Section IV, a system-level simulation of the proposed system is introduced, with an analysis (using 3D ray-tracing) of the antenna\apo s throughput performance in real-life scenarios. Section V concludes our paper.


\section{mmWave Hybrid Beamforming with Lens Antennas}
\subsection{What is Lens Antenna Hybrid Beamforming?}
In general terms, a lens is a refracting device that focuses an incident electromagnetic wave. In wireless communications, a lens can be described as a passive phase shifter that modifies the input signal phase according to its incident point on the lens aperture. Consequently, the lens focuses signal energy to a subset of antennas according to the angle-of-departure (AoD) or angle-of-arrival (AoA); this is called the \textit{angle-dependent energy focusing} property~\cite{RuiZhang2016opdm}. This property contributes to the lens antenna\apo s  high gain and directivity since the beam is focused in a certain direction with concentrated signal power. 

The targeted mmWave lens hybrid beamforming system to which we ultimately aim to apply our proposed lens antenna is shown in Fig.~\ref{fig1:lensbf_archi}(a). While the baseband and transmitter/receiver (T/R) modules\apo~structure follows that of a conventional hybrid beamforming system, for lens antenna beamforming, a lens is positioned in front of the $N$ element antenna array to exploit the properties of the lens. The lens and the antenna array, as a whole, are referred to as the $N$ element lens antenna.
\begin{figure*}[t]
	\centering{\includegraphics[width=1.7\columnwidth,keepaspectratio]{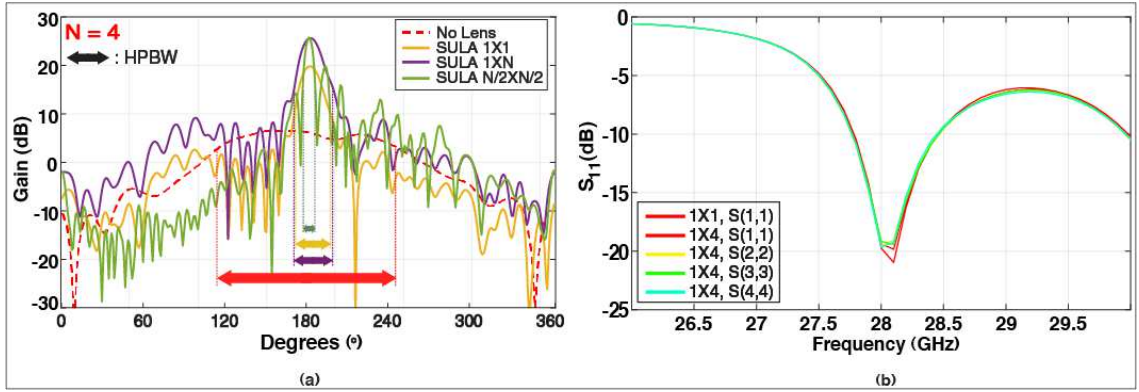}}
	\caption{Measured and simulated antenna radiation pattern and \Sparam s~for the SULA (vertical plane): (a) antenna radiation pattern for the SULA; (b) \Sparam s~for \onebyone and \onebyfour SULA.} \label{fig3:SULA}
\end{figure*}

The \textit{angle-dependent energy focusing} property of lens provides unique benefits to the hybrid beamforming system. First, the system becomes practical and cost-efficient when only a subset of antennas is selected--antenna selection--for signal processing. Depending on the pre-decided AoD or AoA, a controller calculates which subset of the array elements should be switched on to electronically steer the antennas to a specific direction (See Fig.~\ref{fig1:lensbf_archi}(b)). By replacing the heavy network of phase shifters with the switching network with lens, the signal processing complexity and RF chain cost are significantly reduced. Moreover, the high antenna gain of the lens antenna from directivity compensates for the expected performance degradation of antenna selection which is the opposite to when lens is not used~\cite{molisch2004antennaSel}.



\subsection{Feasibility of Lens Antenna Hybrid Beamforming}
Several studies focused on the theoretical and measurement based analysis of actual prototypes of the mmWave lens hybrid beamforming system. In this section, we present previous research on both the theory and measurements of such system. As for theoretical work, the authors in \cite{akbar2013beam} introduced discrete lens arrays (DLAs) and antenna selection-based systems to mmWave hybrid beamforming. The lens acts as an approximate spatial Fourier transformer, projecting signals to the beamspace domain which significantly reduces the number of required RF chains. Multi-path environments and multiuser scenarios were also evaluated~\cite{sayeed2013MUbeamspace}. 

In addition to the theoretical work, several researchers investigated the actual prototyping of mmWave lens hybrid beamforming systems. In~\cite{akbar_2017}, researchers configured a lens array multi-beam MIMO testbed with a Rx lens module with a maximum of 4 RF chains and a single feed open-ended waveguide Tx. For each RF chain, a subset of 4 feeds was assigned, and with four switches, one of the feeds out of the four was chosen from each RF chain to change the beam. The lens was placed in front of the 16-feed array. Under a multi-user MIMO (MU-MIMO) OFDM system setup, results showed that the prototype was able to separate the mixed signals coming from the two Txs. 

The authors in~\cite{Juha_2016} presented a two-dimensional (2D) beam steerable lens antenna prototype operating at 71-\SI{76}{\giga\hertz} with a 64 element feed antenna array. 
Beam-steering and -switching were implemented by simply selecting one of the 64 antenna elements using RF switches integrated into the module. In a near-field antenna test range, the maximum measured directivity and gain were approximately 36.7~dB and 15~dBi, with a beam steering range of $\pm4^\circ\times\pm17^\circ$. A link budget analysis showed 
700~Mbps throughput for a transmission length of 55m.

Previous theoretical research and prototyping studies provide support for the practicality and feasibility of the mmWave lens beamforming system. It is unclear, however, how the actual lens antennas will perform in real environments with real blockages and how the lens antennas\apo~performance will differ based on their design. Moreover, cases in which a lens is not used also have to be evaluated thoroughly to fully understand the usage and properties of a lens antenna in beamforming. Hence, in this article, we evaluate different types and sizes of lens antennas compared to when a lens is not used. We also present their performance through link- and system-level analysis.


\section{\SI{28}{\giga\hertz} Lens Antenna Prototype: Fabrication and Measurements}
\subsection{Lens Antenna Prototype}
The different types of \SI{28}{\giga\hertz} lens antenna prototypes depend on their design. Certain types can perform better than others in specific circumstances. This article evaluates two types of lens antenna prototypes--static-user lens antenna (SULA) and mobile-user lens antenna (MULA). The SULA is for applications where the user is static, so beam-switching is unnecessary and all that is needed for high performance is directivity gain. The MULA is for applications where the user is mobile, so both beam-switching and high directivity gain should be achieved.

The SULA is cube shaped. In the cube, a \twobytwo patch array antenna is placed behind the lens with a polyethylene wall surrounding the gap between the lens and the antenna. For higher array gain, the single SULA can be concatenated into arrays of size \nbyn or \onebyn. The single SULA size is $50 \times 50 \times 60 \mm^3$, small enough for a massive MIMO system. The MULA, in contrast, is not cube shaped, which facilitates beam-switching, as the cube walls can limit the beam-switching angle by converging beams to a certain direction. The lens for the MULA is placed in front of the patch array antenna for a certain focal distance $f$, with four thin polycarbonate cylinder pillars placed on each of the corners for fixation of the lens to the patch array antenna. Detailed descriptions are in Fig.~\ref{fig2:antenna_pic}(a).
\begin{figure*}[t]
	\centering{\includegraphics[width=2\columnwidth,keepaspectratio]{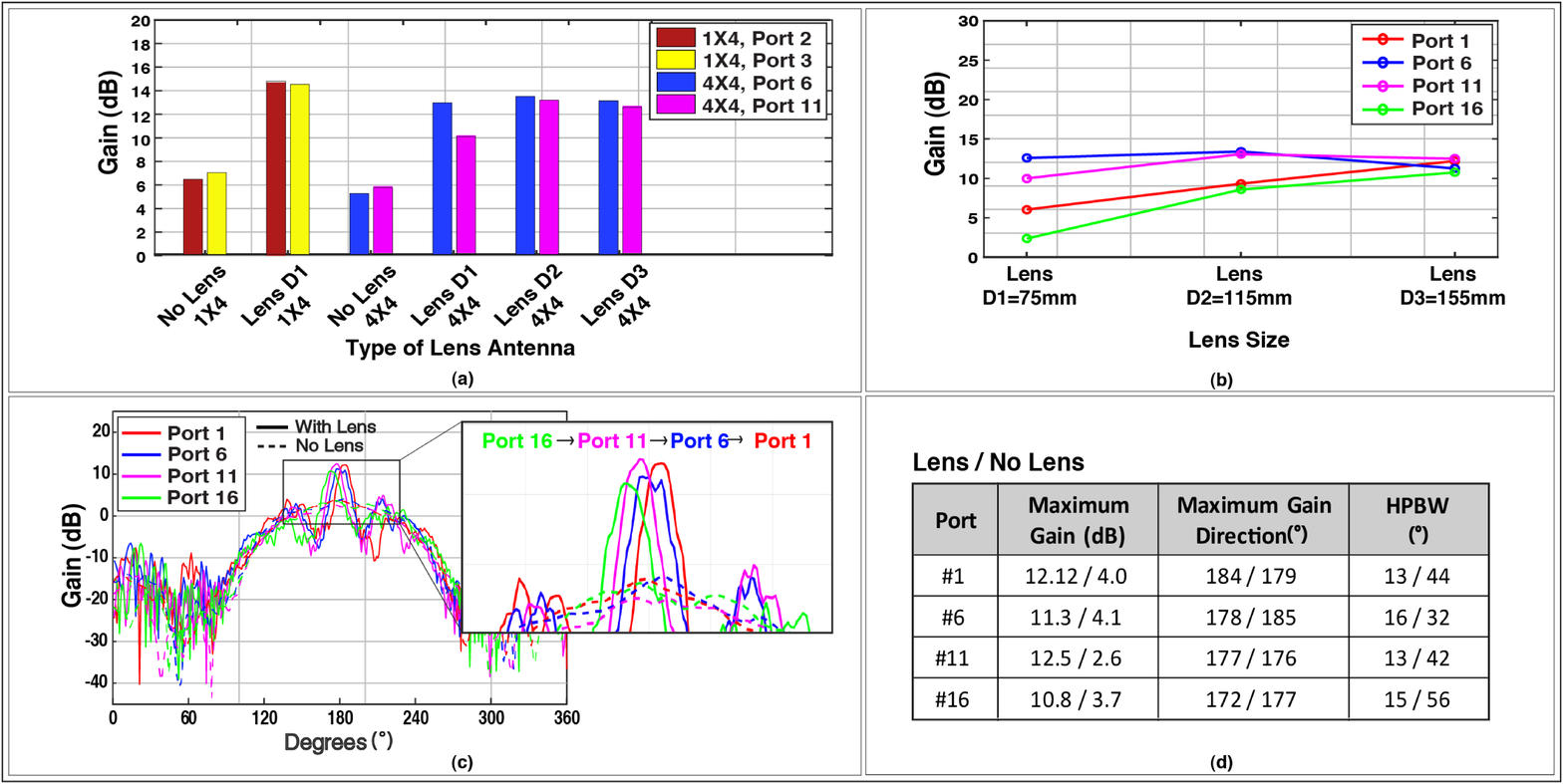}}
	\caption{Measurement based performance analysis of the MULA (vertical plane): (a) maximum achievable gain comparison for differently activated ports of \onebyfour and \fourbyfour MULA; (b) maximum achievable gain comparison of \fourbyfour MULA for differently sized lenses; (c) antenna radiation pattern for \fourbyfour MULA with largest lens ($D_3=\SI{155}{\milli\meter}$) compared to `no lens'; (d) statistics of maximum gain and beamwidth for \fourbyfour MULA compared to `no lens'.} \label{fig:MULA_gain_BSW}
\end{figure*}

For both types we use an identical patch antenna, a square of side length $\ell=3.05\mm$. For an array of patch antennas the inter-element distance between patches is $\lambda\approx10\mm$, where $\lambda$ is the wavelength for \SI{28}{\giga\hertz}. Each patch antenna is a port capable of sending a single beam, enabling beam-switching if multiple ports are each sequentially activated. Each port of the \onebyfour and \fourbyfour MULA is given specific index numbers to facilitate analysis of different scenarios of activated ports in Section~III-B. The details are depicted in Fig.~\ref{fig2:antenna_pic}(b).

The lens used for both types is a hyperbolic, dielectric lens made from polyethylene (dielectric constant $\varepsilon_r=2.40$). The lenses are designed using the well-known principles of classical lenses~\cite{Bahaa1991Photonics,kwon2016rf}. While there is no size variation in the lenses for the SULA, three differently sized lenses with diameters $D_1=\SI{75}{\milli\meter}, D_2=\SI{115}{\milli\meter}, D_3=\SI{155}{\milli\meter}$ are fabricated and evaluated for the MULA (See Fig.~\ref{fig2:antenna_pic}(c)). For a fixed center frequency of \SI{28}{\giga\hertz} and wavelength $\lambda$, the diameter is defined as $D_i=L_{4\times4}+2\alpha_i$ where $\alpha_i=\lambda(2i-1)$ for $i=1,2,3$ and $L_{4\times4}=\SI{55}{\milli\meter}$ is the fixed side length of the \fourbyfour patch array antenna. The ratio of focal length to diameter (i.e., $f_i/D_i$), was fixed to $1.2$ for all cases. 
In this article, we have analyzed lens antenna arrays of sizes up to \twobytwo for the SULA and \fourbyfour for the MULA. However, note that size is not limited and can be modulated to larger arrays depending on required specifications.\footnote{One might argue that \onebyfour and \fourbyfour antenna arrays do not have enough antenna elements to achieve a wide beam scanning angle for beam-switching. The goal of this article is to first demonstrate the feasibility of the lens to be used in the mmWave hybrid beamforming system. Our future plan is to generalize the array up to a 64 x 64 configuration.} 

\subsection{Measurements and Analysis of the \SI{28}{\giga\hertz} Lens Antenna}

In this section, the simulated and measured antenna pattern and gain for the SULA and MULA are presented. We compare our lens prototypes to cases in which no lens is used--`no lens'.\footnote{Note that `no lens' indicates the removal of the lens from any type of lens antenna so that only the remaining patch antenna array is evaluated. For example for SULA, `no lens' will indicate the \twobytwo patch array antenna that is placed underneath the lens. For the \fourbyfour MULA `no lens' will be the \fourbyfour patch array antenna.} For \onebyone SULA, \fourbyfour MULA and `no lens', we present the actual measurements made in an anechoic chamber; the other data presented is simulation data from High Frequency Electromagnetic Field Simulation (HFSS). Moreover, the antenna radiation pattern presented in this article is that of the vertical plane since all antennas are set to vertical polarization.

In Fig.~\ref{fig3:SULA}(a) we present the radiation pattern at vertical plane of `no lens' and the SULAs. Compared to `no lens', the achievable gain of the SULA was 12--17~dB higher with half power beamwidth (HPBW) of $\pm 10^\circ$ on average. While nearly zero directivity and low gain were measured for `no lens', the SULA achieved a distinctive directivity (small HPBW) and gain even for the smallest \onebyone SULA. In addition, if we compare SULA $N/2 \times N/2$ and $1\times N$, we can see that even with the same number of single SULAs, depending on the structure, a narrower beam can be formed while having identical gain. Fig.~\ref{fig3:SULA}(b) presents the measured \Sparam s \ which were given at most -10~dB for 27.8--\SI{28.4}{\giga\hertz} demonstrating that the patch antenna covers well the \SI{28}{\giga\hertz} frequency. 

Secondly, in Fig.~\ref{fig:MULA_gain_BSW}, we analyze the MULA in mainly three aspects--power gain, directivity, and beam-switching feasibility. The crucial property of the MULA to support mobile users is beam-switching implemented by activating a single port amongst multiple ports so that depending on which port is activated, the beams are differentiated in space when going through the lens.

In Fig.~\ref{fig:MULA_gain_BSW}(a), the maximum achievable gains of the different types of MULA and `no lens' are compared for activated ports near the center of the lens. For both \onebyfour and \fourbyfour MULAs, using a lens yielded higher achievable gain for all activated ports of maximum 8~dB. For the \fourbyfour MULA, increasing the size of the lens had a minor effect on the gain which indicates that for the same gain, a smaller lens is a reasonable choice regarding the trade-off between size and gain. To further investigate the effect of lens size, Fig.~\ref{fig:MULA_gain_BSW}(b) illustrates the gain depending on the lens size for four ports--1, 6, 11, and 16 (diagonal order in the \fourbyfour MULA). When the lens was enlarged, the ports near the edge of the lens (ports 1 and 16) benefited more than those near its center (ports 6 and 11). This is reasonable, since the edge ports are more likely to be out of range of the lens focal region, a problem that is solved when the lens gets larger and provides more coverage.
\begin{figure*}[t]
	\centering{\includegraphics[width=1.8\columnwidth,keepaspectratio]
		{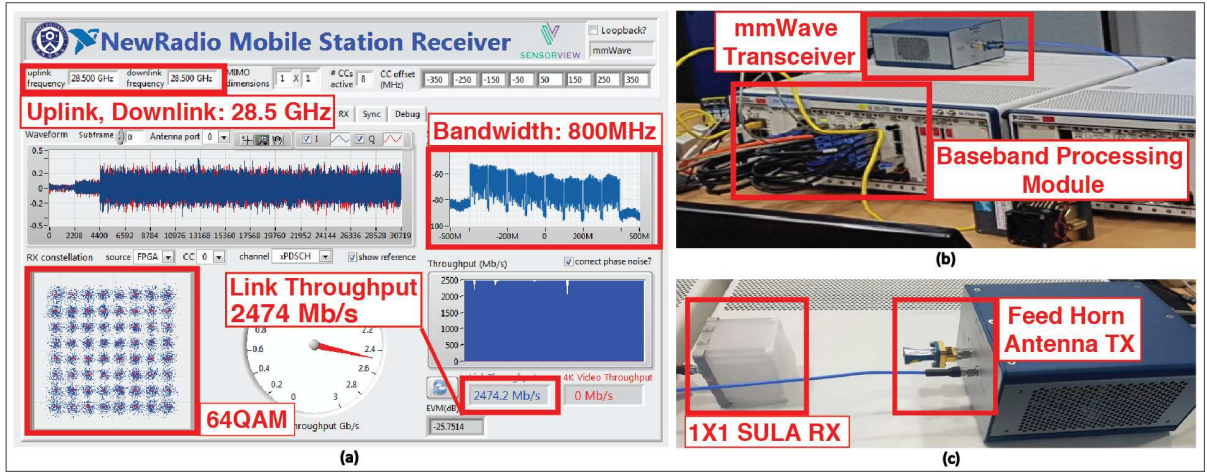}}
	\caption{(a) Link-level performance analysis of \onebyone SULA at \SI{28.5}{\giga\hertz}; (b) system equipment for the link-level analysis; (c) antenna configuration example.} \label{fig5:LLS}
\end{figure*}


Analyzed in Fig.~\ref{fig:MULA_gain_BSW}(c)-(d) are the directivity and beam-switching feasibility of the \fourbyfour MULA. Fig.~\ref{fig:MULA_gain_BSW}(c) depicts the radiation pattern of a \fourbyfour MULA at vertical plane with a lens size of $D_3=155\mm$ and `no lens'. For `no lens', the radiation pattern for all activated ports are nearly identical without any directivity or gain difference. When a lens is used, however, the maximum gain beam direction is shifted to the right in decreasing order of port index (port $16 \rightarrow 11 \rightarrow 6 \rightarrow 1$). For more detailed analysis, we present the statistics for maximum gain, maximum gain beam direction, and HPBW in Fig.~\ref{fig:MULA_gain_BSW}(d). The table shows that the gap in both gain and beamwidth between using a lens and `no lens' is large. The HPBW shows at most $41^\circ$ gap between lens and `no lens' with gain difference of at most 10~dB. 

\begin{figure*}[t]
	\centering{\includegraphics[width=2\columnwidth,keepaspectratio]{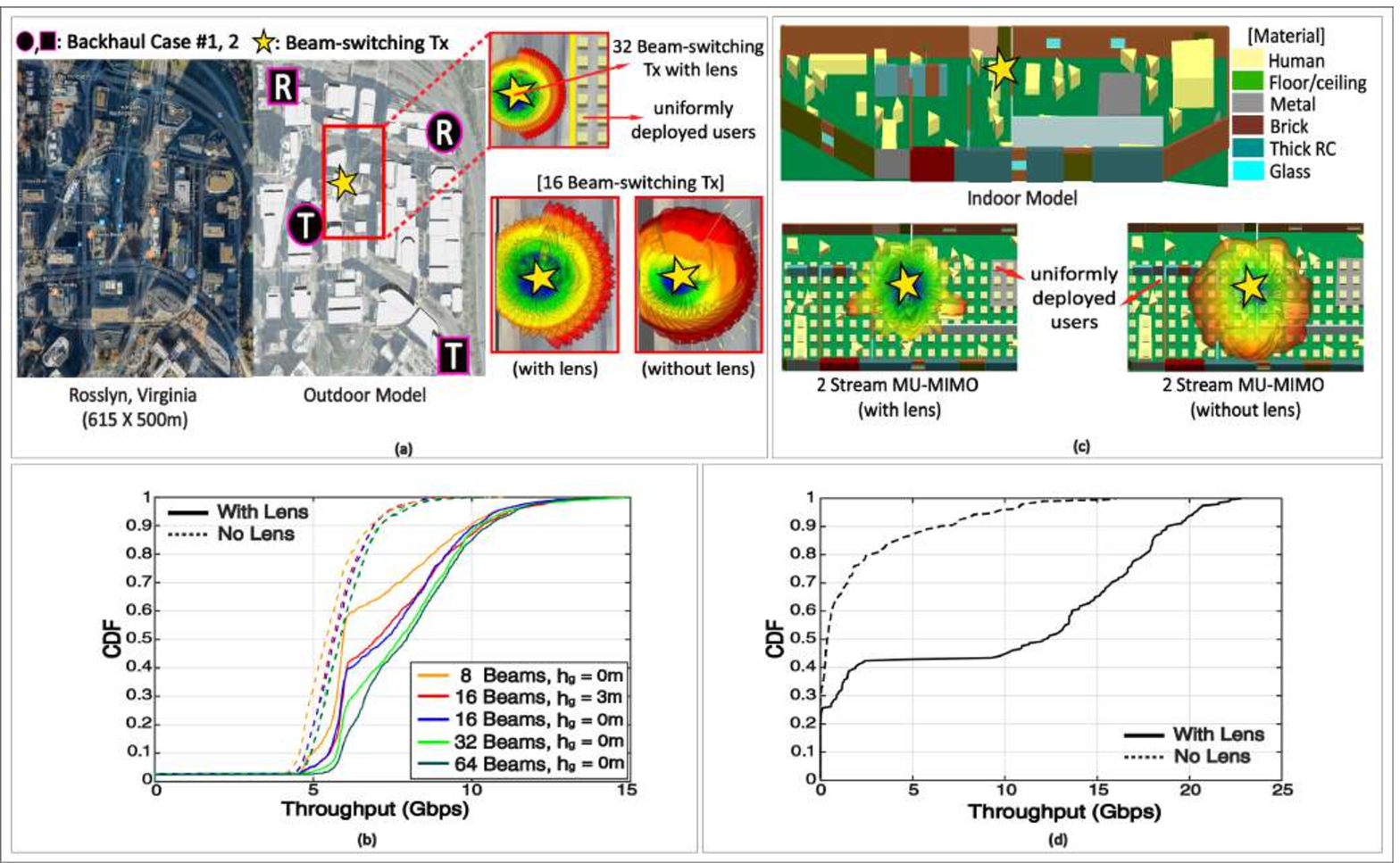}}
	\caption{System level analysis in real environments by 3D ray-tracing: (a) outdoor scenario for backhaul and MU-MIMO beamforming performance analysis; (b) average throughput CDF of outdoor MU-MIMO; (c) indoor scenario for MU-MIMO beamforming performance analysis; (d) average throughput CDF of indoor MU-MIMO.} \label{fig:raytrace_sce}
\end{figure*}

From the analysis above, we can summarize that SULA achieves a high gain of maximum 25~dB--approximately 17~dB higher than `no lens'. The results for gain, directivity, and beam-switching property of the MULA indicate that switching the activated ports in diagonal order shifts the maximum gain direction of the lens antenna with HPBW as small as $\pm6.5^\circ$ and gain as high as 12.5~dB. Such properties of our prototypes strongly support their feasibility for use in the targeted mmWave lens hybrid beamforming system.



To further test our lens antenna, the link-level performance of the \onebyone SULA was evaluated by using a mmWave transceiver SDR setup. Note that the \onebyone SULA is a lower bound for achievable link-level performance of SULA, as it has the lowest gain amongst the SULAs. We used the PXIe (peripheral component interconnect express extensions for instrumentation) platform and multi-FPGA (field programmable gate array) processing to implement the SDR system. With \SI{800}{\mega\hertz} bandwidth, a single data stream is transmitted and received in 64-QAM between a feed horn antenna Tx and a \onebyone SULA Rx~(See Fig.~\ref{fig5:LLS}(a)-(c)) with transmission length of \SI{70}{\centi\meter}. Our results showed a maximum throughput of 2474~Mbps\footnote{Full demo video is available at http://www.cbchae.org/}. 

\section{System-Level Analysis of the Lens Antenna in Real Environments}
Along with the measurement based and link-level analysis of our proposed lens antenna, the system-level analysis in real environments is presented. Using the measurement data from Section III, we carried out a system-level analysis in outdoor and indoor environments using 3D ray-tracing for different scenarios. We considered three scenarios: (i) backhaul usage of SULA in outdoor, (ii) MU-MIMO of MULA with beamforming in outdoor, and (iii) MU-MIMO of MULA with beamforming in indoor. For all scenarios, with \SI{28}{\giga\hertz} center frequency and \SI{2}{\giga\hertz} bandwidth, the lens antennas with the highest gain and directivity from each SULAs and MULAs were used to represent the upper bound of the achievable performance of our proposed lens antenna.

To model the urban outdoor scenarios (i) and (ii), we modeled, as shown in Fig.~\ref{fig:raytrace_sce}(a), an actual region in Rosslyn, Virginia. For the backhaul scenario (i), two cases were considered--\# 1: Closer Tx-Rx distance, \SI{450}{\meter}, with NLoS link and \# 2: Further Tx-Rx distance, \SI{636}{\meter}, with LoS link. In both cases, the \twobytwo SULA was used for both Tx and Rx as a backhaul base station assuming perfect beam alignment with a transmit power of 43~dBm. For case \# 1, the achievable throughput was 16.9~Gbps when lens is used while 3.9~Gbps when not used. For case \#2, the throughput was 34.3~Gbps, 18.8~Gbps respectively for lens and `no lens'. The presence of blockage clearly has a more negative effect on throughput than a longer Tx-Rx distance since case~\#2 has nearly two times the throughput of case \#1 when using a lens. Moreover, using a lens yields significantly higher throughput ($>$10~Gbps) than the `no lens' for both cases.

For the outdoor mobile scenario (ii), we simulated a beam-switching MULA Tx with a total transmit power of 38~dBm (lower than scenario (i) since we considered a road side unit Tx) which is capable of transmitting 8, 16, 32, and 64 beams with HPBW of $\pm10.5^\circ, \pm5^\circ, \pm2.5^\circ,$ and $\pm1.25^\circ$ respectively. The boresight was fixed to approximately $\pm75^\circ$ for all number of beams. Regarding the switch loss which is the power loss and time-delay occurred from the RF switch used for switching beams, we assumed a power amplifier that compensated for the 10--20~dB power loss\cite{Juha_2016} and considered time-delay negligible in our snapshot-like 3D ray tracing scenario since five beams were simultaneously transmitted and time variation was not considered. The Rxs are omnidirectional antennas and uniformly deployed around the Tx inside a rectangular area of \SI{20}{\meter}$\times$\SI{200}{\meter}, with an inter-user distance of \SI{2}{\meter}. Each beam of the Tx has the radiation pattern of the \fourbyfour MULA with the largest lens ($D_3=\SI{155}{\milli\meter}$) and activated port 11. We utilized the measured pattern of the \fourbyfour MULA to generate a virtual $1\times N$ beamforming MULA capable of activating $N$ ports as in Fig.~\ref{fig:raytrace_sce}(a). At every simulation trial, the Tx simultaneously sent, in total, five beams where each beam was the optimal one for each of the randomly chosen five users amongst a total of 1100 users. After 220 trials, the average throughput was derived from the signal-to-interference-ratio (SINR) for `lens' and `no lens'. The effect of the height gap, $h_{g}$, between the Tx and Rx was also evaluated where Rx height was fixed to \SI{3}{\meter} and Tx height was either \SI{3}{\meter} or \SI{6}{\meter}. 

In Fig.~\ref{fig:raytrace_sce}(b), the results showed that using a lens provided a higher maximum throughput of 4~Gbps than `no lens' for the same transmit power. This is because when the beams are well aligned, the lens antenna\apo s sharp beam yields higher gain for each user while lowering the interference signal power. Results also imply that the height gap between Tx and Rx has a minor effect on throughput. 8 beams `with lens' has the least improvement from using the lens compared to the other number of beams. Hence, if the number of beams is too small, using a lens becomes less effective. Moreover, as the number of beams is increased up to 64, the performance improvement slows down indicating that the performance improvement from increasing the number of beams may not be favorable due to the rather high cost in hardware and complexity. This saturation of performance as the number of beams increases implies that the trade-off between hardware costs and achieved gain has to be considered when choosing the reasonable number of beams.

The indoor environment for scenario (iii) was modeled as a $\SI{30}{\meter}\times\SI{10}{\meter}$ virtual indoor space where blockages (including humans), walls, and floors were all updated to reflect the mmWave propagation channel (See Fig.~\ref{fig:raytrace_sce}(c)). In scenario (iii), the Tx of height \SI{3}{\meter} with a total transmit power of 13~dBm sent two data streams with equal power, each directed to the right and left, where the beams\apo~main direction angles were separated at $120^\circ$. The same radiation pattern used for scenario (ii) was used for the Tx. The Rxs are omnidirectional antennas and were uniformly deployed around the entire indoor space with an inter-user distance of \SI{1}{\meter} and a height of \SI{1.5}{\meter}. We evaluated the SINR of all users and calculated the averaged throughput for the `lens' and `no lens'. Fig.~\ref{fig:raytrace_sce}(d) shows results congruent with those obtained at the outdoor MU-MIMO, indicating that using a lens yields higher achievable throughput of more than 10~Gbps on average. 

Given the performance analysis of the proposed lens antenna prototypes presented above, the lens antenna proves to be more than sufficient and feasible to be implemented in a mmWave hybrid beamforming system. For the system-level analysis we built a virtual hybrid beamformer with the MULA using simulation and derived results showing that using the lens antenna with mmWave hybrid beamforming can achieve throughput performance much higher than when a lens is not used in both outdoor and indoor scenarios. Moreover the SULA performance, which can be seen as parallel to the analog beamforming performance of a single data stream, also shows a capacity to yield high throughput.

\section{Conclusion}
In this article we proposed two prototypes of \SI{28}{\giga\hertz} lens antennas for static and mobile usage as future devices to be implemented in a lens-embedded mmWave hybrid beamforming system. Our measurement- and simulation-based analyses offer insight into the profitable properties of lens antennas of much higher gain and directivity compared to those in which no lens is used. The lens antennas\apo~beam-switching feasibility was verified by demonstrating the shift in the beam\apo s main direction when changing the activated port sequentially. The lens antennas\apo~design issues were also considered by analyzing the effect of enlarging the lens size. In addition, we presented link- and system-level performance evaluations that show the high throughput performance of the lens antennas in real indoor and outdoor scenarios. Our future work aims to implement an ultra-fast beam-switchable lens antenna based on an mmWave hybrid beamforming system with low complexity algorithms which exploit mmWave massive MIMO properties~\cite{minsoo2016comp}.

\section*{Acknowledgments}
The authors would like to thank I. Jang from SENSORVIEW for the helpful discussions and suggestions on fabricating the lens antenna prototypes.

\bibliographystyle{IEEEtran} 

\bibliography{LensComMagRefs_rev}

\section*{Biographies}
\begin{IEEEbiography}{Yae Jee Cho} received her B.S. degree from the School of Integrated Technology at Yonsei University in 2016 with high honors. She has been a Ph. D student in the School of Integrated Technology, Yonsei University, in Korea since 2016. Her research interests include lens antennas, millimeter-wave communication, vehicular communication, and molecular communication. 
\end{IEEEbiography}

\begin{IEEEbiography}{Gee-Yong Suk} received his B.S. degree in electrical and electronics engineering from Yonsei University in 2016. Now, he is a graduate student in the School of Integrated Technology, Yonsei University. His research interests include millimeter-wave communications, MIMO communications, 5G networks, and estimation theory.
\end{IEEEbiography}

\begin{IEEEbiography}{Byoungnam~Kim}
received his Ph.D. from Korea Advanced Institute of Science and Technology (KAIST), Korea. He was CTO of Ace Technology and he is now CEO of SensorView Ltd. His research interests include advanced RF technologies for 5G.
\end{IEEEbiography} 

\begin{IEEEbiography}{Dong Ku Kim} received his Ph.D. from the USC, CA, USA in 1992. He worked on CDMA systems in Motorola at Fort Worth, Texas, USA in 1992. He has been a professor in the School of EEE, Yonsei Univ. since 1994. He is a chair of the executive committee of 5G Forum in Korea. He received the Award of Excellence in the leadership of 100 Leading Core Technologies for Korea 2020 from the National Academy of Engineering. 
\end{IEEEbiography}

\begin{IEEEbiography}{Chan-Byoung Chae}(SM'12)
	is the Underwood Distinguished Professor in the School of Integrated Technology, Yonsei University. Before joining Yonsei, he was with Bell Labs, Alcatel-Lucent, Murray Hill, NJ, as Member of Technical Staff, and Harvard University, Cambridge, MA, as Postdoctoral Research Fellow. He received his Ph.D. degree in ECE from The University of Texas at Austin in 2008. He was the recipient/co-recipient of the IEEE INFOCOM Best Demo Award (2015), the IEIE/IEEE Joint Award for Young IT Engineer of the Year (2014), the KICS Haedong Young Scholar Award (2013), the IEEE Sig. Proc. Mag. Best Paper Award (2013), the IEEE ComSoc AP Outstanding Young Researcher Award (2012), and the IEEE Dan. E. Noble Fellowship Award (2008). He currently serves as an Editor for IEEE Trans. on Wireless Comm., the IEEE Comm. Mag., the IEEE Wireless Comm. Letters, and IEEE Trans. on Molecular, Biological, and Multi-scale Comm.
\end{IEEEbiography}

\end{document}